\documentclass[12pt,preprint]{aastex}

\def\Msun{$\rm M_\odot~$}
\def\Lsun{$\rm L_\odot$}
\def\Teff{$T_{\rm eff}$}

\def\simgt{\lower.5ex\hbox{$\; \buildrel > \over \sim \;$}}
\def\simlt{\lower.5ex\hbox{$\; \buildrel < \over \sim \;$}}
\def\ltsima{$\; \buildrel < \over \sim \;$}
\def\gtsima{$\; \buildrel > \over \sim \;$} 
\def\lsim{\lower.5ex\hbox{\ltsima}} 
\def\gsim{\lower.5ex\hbox{\gtsima}} 
 
\def\psr{PSR~J1740-5340~}
\def\pspin{P_{\rm spin}}
\def\pdot{\dot{P}_{\rm spin}}
\def\porb{P_{\rm orb}}
 
\begin{document} 
 
\title{PSR  J~1740-5340: accretion inhibited  by radio--ejection  in a
binary millisecond pulsar in the Globular Cluster NGC 6397}
 
\author{Luciano   Burderi,    \altaffilmark{1}   Francesca   D'Antona,
\altaffilmark{1} Marta Burgay, \altaffilmark{2}}
 
\affil{\altaffilmark{1}Osservatorio Astronomico  di Roma, via Frascati
33, 00127 Roma, Italy}
\affil{\altaffilmark{2}Universit\`a   di   Bologna,  Dipartimento   di
Astronomia, via Ranzani 1, 40127 Bologna, Italy}

\begin{abstract}  
We present an evolutionary scenario for the spin-up and evolution
of binary millisecond pulsars, according  to which the companion of the
pulsar \psr,  recently discovered as  a binary with orbital  period of
32.5 hrs in the Globular Cluster  NGC 6397, is presently in a phase of
`radio--ejection' mass loss from  the system.  The optical counterpart
is a star  as luminous as the cluster turnoff stars,  but with a lower
\Teff\ (a  larger radius) which we  model with a star  of initial mass
compatible  with the  masses evolving  in the  cluster  ($\simeq 0.85$
\Msun). This star has  suffered Roche lobe overflow while  evolving off the
main sequence, spinning up the neutron star to the present
period  of 3.65 ms.  At  present, Roche lobe overflow  due to the
nuclear  evolution of  the pulsar  companion  and to systemic angular
momentum losses by magnetic braking is still going on,
but accretion is inhibited by the momentum exerted by the radiation of the
pulsar on the matter at
the inner Lagrangian point.
The presence of this matter around the system is consistent with
the long lasting irregular radio
eclipses seen in the system. Roche lobe deformation of the mass losing
component is  also necessary to  be compatible with the  optical light
curve.   The   ``radio--ejection"  phase,  which   had  been  recently
postulated  by   us  to  deal  with   the  problem  of   the  lack  of
submillisecond pulsars, can be initiated only if the system is subject
to intermittency in the mass  transfer during the spin--up phase.
In fact, only when the system is detached the pulsar radio emission is
not quenched, and may be able to prevent further mass accretion due to
the action  of the pulsar pressure  at the inner  Lagrangian point.
We propose and discuss that a
plausible reason  why  the  system is  expected  to detach is
the irradiation  of the mass  losing component  by the
X--ray  emission  powered  during  the accretion  phase.   We  finally
discuss the consequences of the  binary evolution leading to \psr, and
its relation with other possible optical counterparts.
\end{abstract} 
 
\section{Introduction} 
\label{sec:intro}  
The widely accepted scenario for the formation of a millisecond pulsar
is the  recycling of an old  neutron star (hereafter NS)  by a spin-up
process  driven by  accretion of  matter and  angular momentum  from a
Keplerian disc, fueled {\it via}  Roche lobe overflow of a binary late
type  companion  (see  Bhattacharya  \&  van den  Heuvel  1991  for  a
review). If  the NS has a  magnetic dipole moment  (typical values are
$\mu \sim 10^{26}\--10^{27}$ G cm$^{3}$)  the disc is truncated at the
magnetosphere,  where   the  disc  pressure  is balanced by
the magnetic pressure  exerted by the NS magnetic  field ($P_{\rm MAG}
\propto  \mu^{2} r^{-6}$), where $r$ is the generic radial distance  
from the NS center.   Once  the accretion  and spin-up  process
ends, the NS is visible as a millisecond radio pulsar (hereafter MSP).
Indeed,  a  common requirement  of  all  the  models of  the  emission
mechanism  from   a  rotating  magnetic  dipole  is   that  the  space
surrounding the NS  is free of matter up to  the light cylinder radius
$R_{\rm LC}$ (at  which the speed of a body  rigidly rotating with the
NS equals the speed of light) .

An interesting evolutionary phase  can occur during the accretion onto
the NS  if the mass transfer  rate drops below the  level required to
allow the expansion of the
magnetosphere beyond $R_{\rm  LC}$ switching--on  the
emission from the rotating magnetic dipole (e.g. Illarionov and
Sunyaev, 1975; Ruderman, Shaham \& Tavani 1989; Stella et al., 1994).
In a standard Shakura Sunyaev accretion disk, it is common to identify four
zones based on the different roles of opacities (free--free or electron
scattering) and pressures (radiation or gas pressure).
The radius of separation between the zone (B), in which the electron
scattering opacity and gas pressure dominate, and the most external zone
(C), where the free-free opacity and gas pressure dominate, is
$
R_{\rm BC} = 4.18 \times 10^{7} L_{37}^{2/3}R_{6}^{2/3}m^{-1/3}
\;{\rm cm}
$\
(see, e.g. Burderi, King \& Szuszkiewicz 1998),
where $L_{37}$ is the luminosity in units of $10^{37}$ erg s$^{-1}$,
$R_{6}$ is the NS radius in units of $10^{6}$ cm, and $m$ is the
NS mass in solar masses. The typical distance of the inner
Lagrangian point from the NS is larger than $\sim 10^{11}$ cm for 
orbital periods longer than $\sim 30$ hr.
Close to the Roche-lobe the disc is then in zone C, where opacity is 
dominated by free-free processes and
pressure is dominated by the gas contribution. In this case, $P_{\rm DISC}
\propto \dot{M}^{17/20} r^{-21/8}$, where $\dot{M}$  is the accretion rate.
The  radial dependence of the pulsar radiation pressure is $P_{\rm RAD}
\propto \mu^{2}  \pspin^{-4} r^{-2}$~ (where $\pspin$~ is the NS  
spin period), and
then it is flatter than the radial dependence of the disc pressure close to
the inner Lagrangian point.
Therefore, $P_{\rm RAD}$ dominates over $P_{\rm DISC}$ for  any $r > R_{\rm
LC}$~ and the whole accretion  disc is swept away  up to the  inner
Lagrangian point $L_1$ by  the radiation  pressure of the  pulsar. During
this ``radio ejection'' phase,  the mechanism that drives {\it  mass
overflow} from $L_1$ can  well be active, but  {\it the pulsar  radiation
pressure at $L_1$ prevents mass accretion onto the NS}.

The occurrence  of a radio--ejection  phase in the  pre--MSP evolution
has been recently proposed by  Burderi et al.  (2001a,b), to deal with
the problem  of the apparent lack of  sub--millisecond pulsars, namely
radio pulsars with spin periods below 1 ms, (Burderi \& D'Amico, 1997)
and with the problem that MSP do never seem to harbor very massive NSs
(see e.g. Taam, King \& Ritter,  2000, Tauris \& Savonije 1999), as it
would be the case if conservative or quasi--conservative evolution has
taken place.

A typical progenitor system of binary  MSPs is made by a NS (hereafter
the primary) accreting from a giant,  or from a  star on its  way to
become a  giant (hereafter the  secondary).  This is testified  by the
many MSPs with low mass ($M  \simeq 0.2 - 0.4$ \Msun) white dwarf (WD)
binary  companions (e.g. Rappaport et al. 1995,  van Kerkwijk  et al.  1996,
Burderi et al. 1998), the remnants predicted by this type of semidetached
evolution.
The NS spin--up occurs {\it  via} the accretion of angular momentum of
the  mass overflowing  the  Roche  lobe. This  mass  transfer is
dictated by the nuclear evolution of the secondary, and the losses of
systemic angular momentum also contribute to determine the
mass transfer rate.  
It is well possible that mass transfer during the spin--up phase
suffers instabilities (see below) and the system temporarily detaches,
allowing  the pulsar to switch--on. However the companion  evolution, 
which leads  to radius  expansion,
will lead again to overflow. 
Mass exchange to the NS  and
spin--up  will  go  on,  until  the pulsar  has been so much spun up that
its radiation pressure at the inner Lagrangian point  is high  enough  to
prevent mass  accretion. Modeling of the binary evolution at the time of
action  of this radio ejection  mechanism is non trivial, and an adequate
study of the physics involved and numerical simulations are needed to
adequately investigate this problem, as it has been done for the case of
``evaporation" of low mass companions of MSPs (Tavani \& Brookshaw 1991,
Banit \& Shaham 1992). The destiny of the system is crucially  dependent on
global mass radius exponent of the mass losing component, on the mass ratio
and on the specific angular momentum of the ejected matter $l$ (see
Ritter, 1995, for general discussion of the stability of mass transfer). We
can foresee cases in which $l$\ is very large with respect to the specific
angular momentum of the system, and a dynamical phase of mass transfer can
occur (Burderi et al.,  2002, in preparation). Here, we will follow only the
case in which $l$\ is the specific angular momentum at the inner lagrangian
point, and the evolution is stationary: the orbit widens and the companion
placidly expands on its nuclear timescale, keeping its contact
with the Roche lobe. 

The main consequence of the occurrence of the radio--ejection phase,
will be that {\it the matter lost  from the  secondary will  not be  
transferred to  the  NS}. 
This phenomenon may look  like the so called ``ablation''  of the companion
irradiated  by the  wind  (composed of  electromagnetic radiation  and
relativistic particles) of a MSP (Tavani 1991).  The key difference is
that,  in the  ablation mechanism,  the process  takes place  once the
companion is detached  (and is evolved into a  low--mass white dwarf),
while radio--ejection takes place {\it during the active phase of
Roche lobe overflow by the  secondary}, and the energy requirements are
much easier  to be met. Moreover,  the energy available to ablate the
companion is different,  being the orbital binding energy  in the
case of radio--ejection  and the  pulsar rotational energy  in the  case of
ablation.   Observationally, in  both cases  we  expect to  be in  the
presence of  a radio MSP  which from time  to time is obscured  by the
matter floating around the system.

In this paper we propose that this postulated radio  ejection phase
has now been detected in the system  containing the  eclipsing
millisecond pulsar  \psr, discovered by  D'Amico et al. (2001a)
in the  globular cluster  NGC6397, 
and identified (Ferraro et al.  2001) with a slightly evolved turnoff
star  in  the sample  studied  by Taylor  et  al.   (2001).

We will  provide convincing evidence that \psr is an
example  of a  system in  the radio--ejection  phase, by  modeling the
evolution of  the possible binary  system progenitor. We  also discuss
why we  should expect that this  type of system may  indeed suffer the
detachments required  in order to initiate  the radio--ejection phase:
in  fact, the  X-ray illumination  of  the secondary  star during  the
accretion phases, which leads to the  NS spin up, is such that the mass
transfer rate is  probably very unsteady.   When: \\  i) as a result
of the accretion of matter and angular momentum, the NS spin period is
so short that,  potentially, the radiation pressure of  a pulsar phase
would  be  high  enough  to   overcome  the  pressure  of  the  matter
overflowing the Roche  lobe, and \\ ii) the  oscillations in $\dot{M}$
are large enough to allow the millisecond pulsar to switch--on, \\ the
radio--ejection phase begins, leading  to the appearance of the system
as it looks now.

We  also  show  that the  most  luminous  `hot'  objects in  NGC  6397
identified  by Taylor et  al. (2001)  as low  mass helium  WDs, indeed
follow  the   evolution  of  binaries   which  may  account   for  the
evolutionary status  of \psr,  while the less  luminous stars  in this
sample might have radii too large to be helium WDs.

\section{\psr}
The  millisecond radio  pulsar  \psr was  discovered  in the  globular
cluster NGC 6397  (D'Amico et al.  2001a). It  has the longest orbital
period ($P_{\rm  orb} \simeq 32.5$  hrs) and the most  massive minimum
companion mass (0.18 \Msun) among the 10 eclipsing pulsars detected up
to now.   The spin period ($\pspin  \simeq 3.65 \times 10^{-3}$  s) and its
derivative  ($\pdot =  1.59 \times  10^{-19}$), recently  derived by
D'Amico et  al. (2001b), allow  the determination of the  NS magnetic
moment $\mu_{26} \simeq 7.7$.  Because of its position with respect to
the cluster center,  the contamination of $\pdot$ because  of the NS
acceleration in  the gravitational field of the  cluster is negligible
(D'Amico et al. 2001b), implying  that the estimate of the NS magnetic
moment is  reliable.

Ferraro et al. (2001)
have extracted, from archive HST data,  a star whose position is consistent
with the  position of  \psr (derived  from one year  of pulse
timing by D'Amico  et al.  2001b). This optical counterpart shows  
light modulation at
the same orbital period as the radio data.
D'Amico et al.
(2001b) have studied the radio eclipses of  \psr, which last for  about
40\% of the
orbital phase at  1.4 GHz.  Out of eclipse  the pulsar signal at  1.4
GHz shows significant  excess propagation delays  (up to $\sim 8$  ms) and
strong intensity  variations.  Similar variations  have been observed, at 400
MHz, in the eclipsing  pulsar PSR B1957+20 close to the eclipse ingress and
egress  (Fruchter et al.  1990) and  in PSR J2051-0827, at 1.4 GHz,  when the
pulsed  signal is occasionally detected  during the eclipse (Stappers  et al.
1996).  These similarities  suggest that in \psr the  signal is propagating
through a dense  material surrounding the system.  In order to  investigate
this possibility, D'Amico et al. (2001b) have fitted  the excess delays
measured in  two adjacent bands of 128 MHz each at 1.4  GHz. They found that
the excess delays $\Delta t$  can   be  well  fitted   with  the  equation
$\Delta   t  \propto \nu^{-2.02\pm0.30}$  that strongly  supports the
hypothesis  that the responsible mechanism is dispersion in a ionized medium.
In this case the  corresponding  electron  column  density variations  are
$\Delta n_{\rm  e}  \sim 8  \times  10^{17}  \Delta  t_{-3}$ cm$^{-2}$,
where $\Delta t_{-3}$ is the delay at 1.4 GHz in ms. For $\Delta t_{-3} \sim
8$ the estimated electron column  density is $\sim 6.4 \times 10^{18}$
cm$^{-2}$.

The  eclipsing radius is $R_{\rm E}  \sim a \sin (0.4 \pi)$, as the eclipse
lasts for $\sim$40\% of the orbit
(D'Amico et  al., 2001b). It is $4.4 \times  10^{11}$ cm, if we take
$m_1=1.8$\Msun~  for the NS mass and $m_2=0.45$\Msun~ for the secondary mass
(see below) and $P_{\rm  orb} \sim 32.5$ hr ($a$  is the orbital separation).
This radius is  larger than the Roche lobe radius of the  secondary ($\sim
1.3 \times 10^{11}$ cm). This means that the eclipsing  matter is beyond the
gravitational influence of the  companion star and  must be continuously
replenished. We can estimate a rough order of magnitude of the necessary
mass loss rate from the secondary,
$\dot{M}$, by assuming spherical symmetry (which {\it is not} consistent
with the randomly variable signal intensity shown by the radio data). We
further take the minimum electron column density which, for a system
seen edge--on, occurs  when the  NS is  in front  of the secondary, so that:

\begin{equation}
n_e \ga \gamma
\int_{a   +  R_{\rm  RL1}}^{\infty}   \rho [(X+0.5Y)/m_{\rm
p}]~dr
\end{equation}

where $\gamma$ is the  fraction of ionized gas, $R_{\rm RL1}$
is the radius of the primary Roche lobe, $X \sim 0.7$ and $Y \sim 0.3$
are the hydrogen and helium  mass fractions respectively, 
and $m_{\rm p}$ is
the proton mass. Using for the density $\rho$~
the continuity equation:
\begin{equation}
4 \pi r^2  \rho v =  \dot{M}
\end{equation}
where $v_{\rm  wind}$~ is the speed of the outflowing matter,
a simple integration gives
\begin{eqnarray}
\dot{M}_{-10} \la 1.17 \times n_{20}~ \gamma^{-1} (X+0.5Y)^{-1}~ v_8~
(m_1+m_2)^{1/3} P_{\rm orb \; h}^{2/3} \nonumber \\
\left[ 2 - 0.462 (\frac{m_2}{m_1+m_2})^{1/3} \right]
\label{eq:mdot}
\end{eqnarray}
where $\dot{M}_{-10}$ is the mass  loss rate of the secondary in units
of $10^{-10}$ \Msun yr$^{-1}$, $n_{20}$ is the electron column density
in units of  $10^{20}$ cm$^{-2}$, $v_8$ is the speed in units of
$10^8$ cm  s$^{-1}$, and $P_{\rm orb \; h}$  is the orbital  period in hours.
We adopt the parameters given above, the derived $n_H=6.4 \times 10^{18}$cm
$^{-2}$~ and $\gamma = 1$. We finally choose $v_8=1$,
typical of winds driven by the radiation of a pulsar (Tavani \& Brookshaw
1991) and $\sim$3 times the escape velocity from the companion.
We get $\dot{M}_{-10} \la 1.5$. Even considering the uncertainty on this
estimate,
winds induced by the pulsar radiation are more typically 2--3 orders of
magnitude weaker (Tavani \& Brookshaw, 1991). On the other hand, if such a
wind were independent from the association with the pulsar in a binary, it
would be a common feature in the evolution of the {\it single} stars in NGC
6397. But the survival of lithium at the stellar surface is consistent only
with rates smaller than  $10^{-12}$ \Msun yr$^{-1}$~ (Vauclair and Charbonnel
1995) as a stronger wind, acting over the 10 Gyr of the cluster lifetime,
would expose the  stellar layers  in which  lithium has been destroyed  by
nuclear burning. The turnoff stars in  NGC 6397 {\it do show} lithium at a
level  about normal for population II  stars (Pasquini and Molaro 1996),
putting strong upper limits to the mass loss rates of main sequence winds.

Therefore, the  mass loss rates which we derive are much more consistent with
Roche lobe overflow driven by nuclear evolution of the secondary and orbital
angular momentum mass loss, than with a possible wind from the secondary,
suggested by D'Amico et al. (2001) and Ferraro et al. (2001).
As our model requires that the mass lost is swept out by the pulsar radiation, 
the value of $v_8$\ given above is consistent with our hypothesis.

\section{Binary evolution in the Globular Cluster NGC 6397} 
\label{sec:cluster}
\subsection{The HR diagram}
\begin{figure}
\plotone{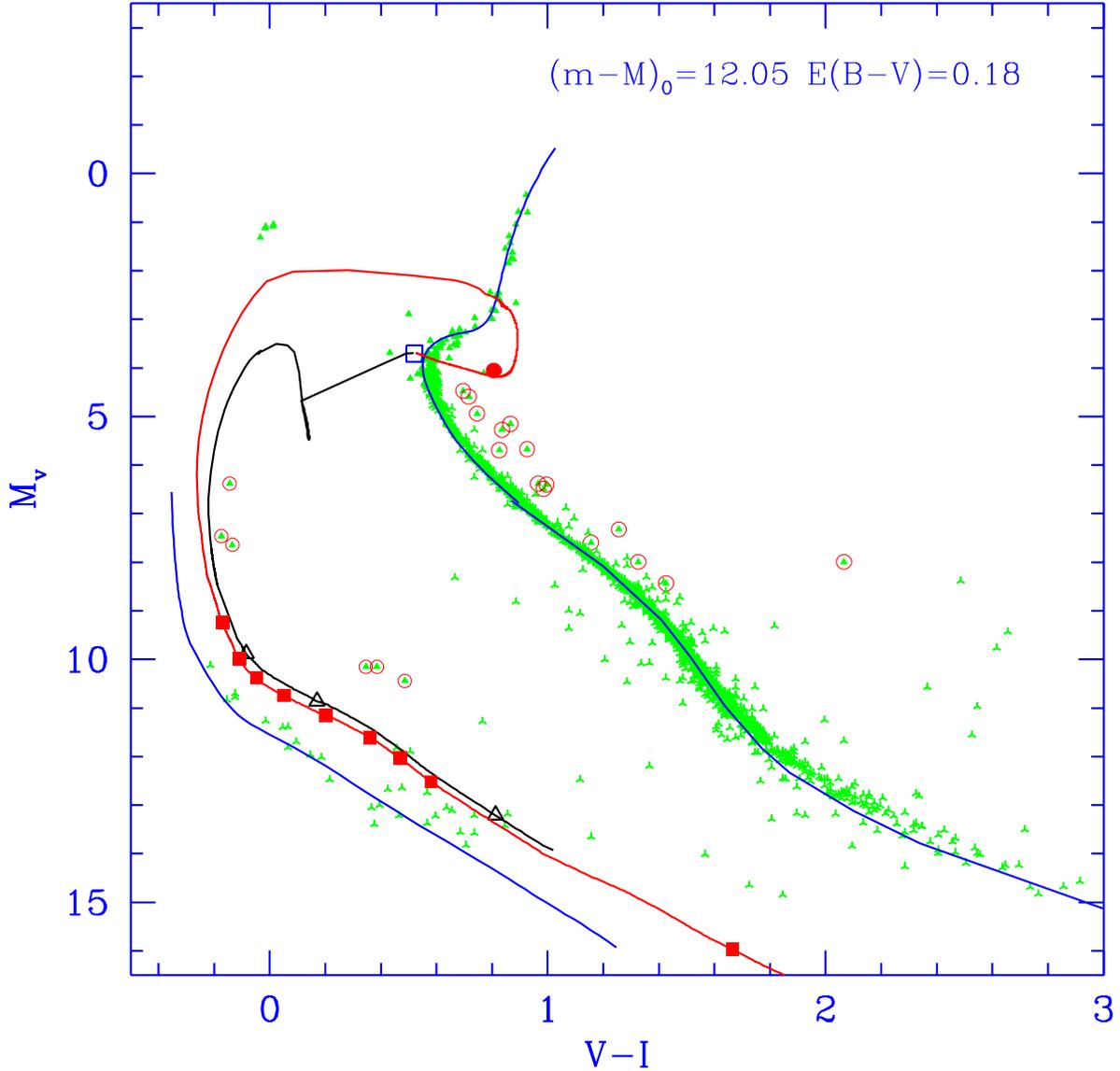}
\caption{\label{fig:fig1} HR diagram of the Globular Cluster NGC 6397, with
an  isochrone superimposed (see  the text for  details). Starting close  to
the  turnoff at  the  open  square,  we show a standard evolution (case
2) of the companion  of the NS, having an initial mass of  0.85\Msun and  an
initial  orbital period  of  14.4 hrs.   (track evolving first  towards the
right  of the figure, and  passing through the observational counterpart of
\psr  -- full dot), and the evolution of case 3, which includes a fixed
irradiation luminosity (track to the left).   Both sequences end  into the
WD evolution  of a  0.246 \Msun (case 2) and 0.216 \Msun (case 3). Both
sequences are followed along the WD cooling,  which  is dominated  by
residual proton  proton burning lasting for more than a Hubble time. Ages of
17, 20 and 26 Gyr are labelled as triangles along the cooling track of case 3.
Ages from 13 to 20 vin steps of 1 Gyr, plus a last point at 25 Gyr, are 
labelled as squares for case 2} \end{figure}

In order  to choose coherently  the input parameters for  the possible
evolution leading to  \psr, it is important to  take into account what
we know of the general properties  of the host cluster, which has been
carefully studied  down to the low  end of the  main sequence.  Figure
\ref{fig:fig1} shows the composite HR diagram of NGC 6397 in the plane
$M_v$ versus $V-I$.  We have plotted the sample by  King et al. (1998)
including the low mass main  sequence (MS) and the white dwarfs (WDs),
together  with the sample  by Cool  (1997) for  the turnoff  and giant
stars.  The  open circles identify the objects  examined by Taylor  et al.
(2001) in the core of this  cluster, to select objects which have been
probably subject to binary evolution.  In fact they recognize a helium
WD sequence on the left of the MS, and many BY Dra stars on its right.
The offset  from the MS suggests  indeed that these  latter stars have
suffered  binary evolution. One  of the  Taylor et  al. (2001)  BY Dra
candidates, plotted  as a  full dot in  our figure  \ref{fig:fig1}, is
indeed the  optical counterpart of  \psr, as discovered by  Ferraro et
al. (2001). On the observational HR diagram we show an isochrone of 12
Gyr for metallicity in mass  fraction Z=0.006 and helium mass fraction
Y=0.23 from  Silvestri et al.   (1998), complemented by the  models by
Baraffe et al.  (1997) for masses $\leq 0.5$ \Msun, while the track on
the cluster white  dwarfs is the 0.5 \Msun  Carbon Oxygen WD evolution
by Wood (1995)  as used in Richer et al. (1997).   The isochrone of 12
Gyr which  fits the cluster  HR diagram implies  that a mass  of $\sim
0.81$ \Msun is  evolving at the cluster turnoff (TO),  and that its TO
luminosity is  $\simeq 2.24$ \Lsun.  Different  interpretations of the
HR diagram  morphology, assuming that  the distance of the  cluster is
smaller,  may lead  to values  of the  TO luminosity  down  $\sim 1.8$
\Lsun.

\subsection{Choice of the initial parameters}
In order  to choose the initial  system parameters which  leads to the
present stage of \psr, we make the following considerations: \\
1) the location of the optical  counterpart is at a luminosity similar
to  the   TO  luminosity  (1.8-2.3   \Lsun,  based  on   the  previous
discussion), but cooler than the TO,  that is, at a radius larger than
the TO radii. The pulsar luminosity is in fact estimated to be $L_{\rm
PSR} = 1.29 \times 10^{35}$ erg s$^{-1}$ (see below), of which, in the
hypothesis of  isotropic emission, a fraction  $R_2^2/(2a)^2$ impacts at
the mass  losing star  surface ($R_2$  and $a$ being  the radius  of the
secondary and the orbital  separation respectively).  Even in the most
favorable case of  a secondary filling its Roche  lobe the fraction of
the  pulsar  luminosity  impacting   the  secondary  is  $f  =  R_{\rm
RL2}^2/(2a)^2  =  5.34  \times 10^{-2}  [m_2/(m_1+m_2)]^{2/3}$,  where
$R_{\rm  RL2}$  is  the  Roche  lobe  radius  of  the  secondary.   We
considered secondary  masses ranging  from 0.25 up  to 0.81  \Msun and
corresponding NS masses from 2.0  down to 1.4 \Msun (conservative mass
transfer). We obtained  values of $f \times L_{\rm  PSR}$ ranging from
$0.4$  to $0.9$  \Lsun respectively.   Therefore the  fraction  of the
pulsar luminosity  reprocessed by the secondary is  not very important
for  determining the  HR diagram  location  of the  star. The  orbital
period is  moderately large, so any  stage of mass  transfer must have
begun while the mass losing component was still not much evolved. \\
2) We exclude the possibility that the star is a normal MS star losing 
a stellar wind at
rates $\sim 10^{-10}$\Msun/yr, for the reasons given at the end of Sect.
2.\\
3) The pulsar spin up to 3.65 ms  must be due to mass exchange in
a previous evolutionary  phase. 
Indeed the possibility that the NS was born with such a period 
seems unlikely as the spin-down age of \psr is $\tau = \pspin/2\pdot 
\sim 350$ Myr (D' Amico et al. 2001), which is much shorter than the
age of the globular cluster NGC 6397 in which \psr is located.
The easiest way is  to attribute it to
the previous evolution  of the system. Therefore we  model the initial
parameter of the system as starting  with a 0.85 \Msun component and a
NS component of 1.4\Msun.
NGC 6397 is a post core collapse globular cluster (Diorgovski \& King 1986),
in which a significant number of interacting neutron star plus main sequence
binaries may have formed by tidal capture or exchange collisions (e.g.
Davies, Benz and Hill 1992, Di Stefano and Rappaport 1993). Implicitly, we
are making the assumption  that the 0.85\Msun\ component has been captured by
the NS at such a separation to be able to begin mass transfer when it is not
yet evolved as a giant. 
This is in line with what has been recently
suggested by Podsiadlowski, Rappaport, and Pfahl (2002 CHECK!) 
that suggest, supported by the numerous ultracompact binaries
found in globular clusters, that tidal capture seem to be a more likely
way to produce this kind of system.

\subsection{The binary evolution}
We  follow  the binary  evolution  with  the  ATON1.2 code  (D'Antona,
Mazzitelli, \& Ritter 1989). The  mass loss rate is computed following
the formulation  by Ritter (1988),  as an exponential function  of the
distance of  the stellar  radius to  the Roche lobe,  in units  of the
pressure scale  height. This method  also allows to compute  the first
phases of  mass transfer, during  which the rate reaches  values which
can  be much larger  than the  stationary values,  due to  the thermal
response of the star to mass  loss. The evolution of the system also
includes orbital angular momentum losses through magnetic braking, in the
Verbunt and Zwaan (1981) formulation, in which the braking parameter is
set to $f=1$. We also tested a case in which $f=2$.

We study four cases of evolution (see Table 1). In all cases, we assume
that the secondary initial mass is $M_2=0.85$\Msun, the primary neutron star
has $M_1=1.4$\Msun, and the orbital initial period is 14.27hr.
The nuclear age of the 0.85\Msun is $\sim 10$Gyr when it begins the Roche
lobe overflow, so that it is only slightly evolved, and hydrogen is almost
exhausted in its (radiative) core. In case 1, we assume a `standard'
conservative evolution. This produces mass  transfer rates of $\sim 2-4
\times 10^{-10}$ \Msun yr$^{-1}$. 
The system  would thus appear as a low mass X--ray
binary  (LMXB).  The  X--ray emission expected  from the  NS is $\sim 2
\times 10^{36}$ erg/s.
The evolution is typical for such systems (e.g. Bhattacharya  \&  van den
Heuvel, 1991). The end product will be a very low mass helium white dwarf
($M_2=0.246$\Msun) in a much wider orbit ($\porb = 119$ hr) 
with a 2\Msun\ neutron
star.
Being the hydrogen not yet completely exhausted in the stellar core
when mass transfer begins, the initial binary period we have chosen
is indeed very close to the ``bifurcation'' period below which
the orbital evolution proceeds towards shorter binary periods.
Note that in our case the bifurcation period is shorter than 
that quoted by Podsiadlowski, Rappaport, and Pfahl (2002) 
for a binary composed by a 1.4 \Msun\ NS plus a 1.0 \Msun\ secondary,
namely $P_{\rm orb \; bif} \sim 18$ hr. This is due to the fact that the main 
sequence radius of our 0.85 \Msun\, population II secondary is $\sim 20\%$
smaller than the radius of a population I, 1.0 \Msun\ companion for
comparable consumption of the hydrogen in the core. Since for a Roche-lobe
filling secondary we have that $\Delta P_{\rm orb}/P_{\rm orb} =
2.2 \Delta R_2/R_2$, the bifurcation
period for the population I, 1.0 \Msun\ companion will be $\sim 44\%$
longer than in our case and therefore consistent with the result of
Podsiadlowski, Rappaport, and Pfahl (2002). 

Having in mind the oberved system \psr, we followed another case of evolution
(case 2) in which we assumed conservative evolution up to an orbital 
period of $\simeq 26$ hr, 
and then we assumed that the matter lost from the secondary is all
lost from the system at the inner lagrangian point L1 with its specific
angular momentum. This alters slightly the conservative evolution, leading to
marginally different mass loss rates and final parameters for the system when
all the seconday hydrogen envelope is lost and the star becomes a white dwarf
(see Table 1).

When mass accretion on the NS is assumed, and the binary evolution suffers a
LMXB phase, it is also important to consider how the X--ray phase would
affect the binary evolution we are considering, as a fraction $R_2^2/(2a)^2 \la
2.7 \times 10^{-2}$ of the X--ray luminosity (see point 1  above) impacts at
the secondary surface. There have been many attempts to model the effect of
this irradiation on the binary evolution, when it has been realized that
spherically symmetric illumination in LMXBs can have a dramatic effect on the
structure of low mass secondaries having convective envelopes. In fact,
X--rays block the intrinsic stellar flux produced by the nuclear reactions,
leading to a stellar expansion, which affects the mass transfer rates
(Podsiadlowski, 1991; Harpaz \& Rappaport, 1991). In reality, illumination
will be non-spherical, because only one hemisphere is affected
-- in fact the donor star is tidally locked
as it is filling its Roche--lobe during this evolutionary phase --, 
and its effect will depend on the depth at which the energy is 
deposited below the
photosphere and how fast it can be transported to the cool side of the star.
If the circulation time is shorter than the cooling time of the heated
matter, we go back to the spherical illumination case. Otherwhise, the global
secular evolution is not much affected, but the mass transfer occurs in
``outbursts" whose duration and peak mass transfer rates depend on the
efficiency of circulation (e.g. Hameury et al. 1993, Harpaz and Rappaport
1994, Vilhu et al. 1994, D'Antona 1994). All the existing computations refer
to the evolution of LMXBs with main sequence companions. We are now dealing
with a case of evolution in which the mass losing component is more evolved,
and in addition has no important convective envelope at the time the mass
transfer begins, so that it is important to have an idea of the possible
effects of X--ray illumination in our specific case.

On the other hand, self--consistent modelling is really very
difficult, as the feedbacks  are not easy  to describe both  physically
and  numerically. However we follow the evolution of an ideal
limiting case, considering a system in which the secondary is affected by a
fixed value  of heating luminosity $L_{h}=5  \times 10^{34}$ erg
s$^{-1}$ (case 3).  We include the effect of illumination as described in
D'Antona and Ergma (1993), following Tout et al. (1989). In this simplified
schematization, the star is immersed in the X--ray radiation bath, and the
total luminosity $L_{tot}$\ which it must radiate is the sum of the stellar
luminosity $L_*$\ plus the heating luminosity $L_h$, from which an
`irradiation temperature' $T_{irr}=(L_h/4 \pi \sigma R_2^2)^{1/2}$\ is defined.
The stellar luminosity and \Teff\ then are related by:

\begin{equation}
L= L_{tot}-L_h=4 \pi \sigma R_2^2(T_{\rm eff}^4-T_{irr}^4)
\end{equation}

Consequently, the stellar \Teff\  becomes hotter due to  the irradiation, and
the star rapidly evolves in the HR diagram at a location determined by the
amount of  irradiation allowed.  The  phases  of  mass transfer  are
slightly altered by  the new system conditions, but  are globally very
similar to case 1. However, as the star loses mass, its radius becomes larger
than the  radius  of the  standard  sequence. This  difference amounts to
$\sim 20$\%  at the orbital  period of \psr. 
Of course the X--ray
irradiation is not self consistently described by this model.  In fact, any
fluctuation in the mass  transfer  rate will  be  amplified. If the mass
transfer decreases further, 
irradiation decreases too, the radius also will decrease,
trying to reach its non  irradiated equilibrium value and the mass transfer
rate decreases more.  We can foresee that the LMXB phase in fact
can alternate phases of mass transfer and detached phases, as it was
predicted for the LMXB having MS secondaries (Hameury et al.
1993, D'Antona 1994, Harpaz and Rappaport 1995).

To test the sensitivity of the mass loss rate to the assumptions made
concerning
magnetic braking, we follow another evolution (case 4) in which $f=2$~ and
the mass transfer is non conservative for $\porb > 17.36$ hr.

\subsection{The final phases of evolution}

Figure 1 shows the evolutionary paths of case 2 and 3 in the HR diagram.
The theoretical values of luminosity and \Teff\ 
are converted into the observed
magnitudes V and I by means described in Bessell, Castelli and Plez (1998).
The irradiated track (case 3) represents the evolution of the
accreting progenitor of \psr and therefore does not reproduce the location 
of \psr for which the accretion is inhibited by the pulsar radiation.
The track passing through the optical counterpart of \psr is the case 2
evolution. The track corresponding to case 1 (purely conservative evoluton)
is very similar, and is not shown.

All the sequences are evolved until the mass loss phase finally ends with the
stellar remnant evolving into the white dwarf region as low mass helium white
dwarfs.
We see that the most luminous three  objects among those identified by
Taylor et al. (2001) as helium WDs  actually may be the end--products of such
an evolution. Notice however that further study is necessary to assess
whether  the lower  luminosity objects  are helium  white  dwarfs.  In
fact, unless their colors are peculiar, they seem too cool to have the
radius expected, even for the lowest possible helium WD masses remnant
of mass exchange  evolution, which are actually $\sim  0.2$ \Msun
(D'Antona et al. 2002, in preparation).
The open triangles show the location  at which sequence 3 achieves a
total age of 17, 20  and 26 Gyr ($\simeq 10.5$ Gyr of  which have been spent
in the phases previous to mass  exchange).  The point at 20 Gyr, which
corresponds more  closely to  the luminosity of  the lowest  Taylor et
al. objects, is  at \Teff = 8900 K. The full squares along sequence 2
correspond to ages from 13 to 20Gyr in steps of 1Gyr, and a last square
indicates the location at 25Gyr.
We also notice  that the cooling of such   objects  (also  in   the  present
models)  is   dominated  by proton-proton  burning of the  remnant hydrogen
layer (Driebe  et al. 1998, Sarna et al.  1998, Sch\"onberner et al. 2000),
and that the low initial metallicity of the system prevents hydrogen shell
flashes.

\section{The evolution of the progenitor system of \psr}

The computation of case 3 has shown that the binary evolution which we are
describing is not dramatically altered by the irradiation due to the LMXB
phase: in particular, the final evolution is very similar, although the
white dwarf remnant mass is slightly smaller. However, it helps to predict
that the LMXB phases can be alternated with phases in which the system
remains detached, as already had been suggested for systems having MS
companions. Let us now consider the evolution of the system, taking also
into account the pulsar's behavior.
Any time the system detaches, the radio pulsar will switch on, but it
will be  quenched  again  when  the  mass transfer  resumes. In the HR
diagram the position of the secondary component will shift from its
`irradiated' position during the LMXB mass transfer phase (track 3 on the
left of the MS) to its `standard' position along track 1.
However,  as
discussed  in Burderi  et al. (2001a,b), when the pulsar is spinning
sufficiently  fast  and  the   orbital  period  is  sufficiently  long
(i.e. the orbital separation is large), the radiation pressure exerted
by the pulsar  at the inner inner Lagrangian point  is larger than the
pressure exerted by the matter  overflowing the Roche lobe even if the
mass transfer rate recovers its  secular value dictated by the nuclear
evolution of  the companion.

This  condition is verified  for orbital periods longer than
%
%
\footnote{
Equation \ref{eq:pcrit} has been derived in Burderi  et al. (2001a,b) 
equating the pressure of a Shakura--Sunyaev accretion disc with the 
radiation pressure exerted by the pulsar. Once the disc has been swept
away the pressure of the flow overflowing the inner Lagrangian point
should be used instead of the pressure of the disc. However it is possible
to demonstrate that these pressures are comparable, which supports the
validity of equation \ref{eq:pcrit}. Let us suppose that the flow of matter 
overflowing the inner Lagrangian point has a cross sectional area 
$A_{\rm flow} = \pi (\delta R_{\rm RL1})^2$ with $\delta \la 0.1$. Moreover
the inward speed of the flow is $v_{\rm flow} = \beta v_{\rm ff}$ where
$v_{\rm ff}$ is the free fall speed and $\beta \la 1$ since the Coriolis
force deviates the flow motion from the radial direction. 
The ram pressure of the flow is 
$P_{\rm flow} = \rho_{\rm flow} v_{\rm flow}^2$, where $\rho_{\rm flow}$
is the density of the flow that is related by the continuity
equation to the mass loss rate. 
The pressure of a Shakura--Sunyaev accretion disc is 
$P_{\rm SS} = \rho_{\rm SS} k T /(n m_{\rm p})$, where $\rho_{\rm SS}$
is the density of the disc that is, also in this case, related by the 
continuity equation to the mass loss rate, $k$ is the Boltzmann
constant, and $T$ is the temperature of the disc at its midplane
(see e.g. Frank, King, and Raine 1992). After some algebraic manipulation,
using the standard relations for a Sakura--Sunyaev accretion disc,
we get $P_{\rm flow} = P_{\rm SS} \times 100 (h/r) (4/\sqrt{2}) \alpha
(\beta/\delta_{0.1}^2)$, where $h$ is the half thickness of the disc
and $\delta_{0.1} = \delta/0.1$. As a good approximation we can take
$h/r \sim 1.7 \times 10^{-2}$ as the dependencies on the disc parameters
are rather weak (see again Frank, King, and Raine 1992). 
Therefore we have 
$P_{\rm flow} \sim 5 \alpha (\beta/\delta_{0.1}^2) P_{\rm SS}$. 
}
%
%
\begin{eqnarray}
P_{\rm orb \; crit} =  0.75~ \alpha^{-54/25}~ n_{0.615}^{-12/5}~ 
m_1^{107/50}~ (m_1+m_2)^{-1/2}~  
\nonumber \\
g(m_1,m_2)^{-3/2}~ \mu_{26}^{-24/5}~ P_{\rm spin \; -3}^{48/5}~ 
\dot{M}_{-10}^{51/25} \; {\rm h}  
\label{eq:pcrit}
\end{eqnarray}
where $\alpha$ is the Shakura--Sunyaev viscosity parameter, $n_{0.615}
= n/0.615  \sim 1$ for  a gas with  solar abundances ($n$ is  the mean
particle mass in units of  the proton mass $m_{\rm p}$), $g(m_1,m_2) =
1 -  0.462 \left( {m_2} \over {m_1+m_2}  \right)^{1/3}$, $\mu_{26}$ is
the magnetic moment  of the NS in units of $10^{26}$  G cm$^3$ ($\mu =
B_{\rm s} R^3$ with $R$ and $B_{\rm s}$ NS radius and surface magnetic
field along the  magnetic axis, respectively), and $P_{\rm spin \; -3}$  
is the NS
spin period  in milliseconds. 
If  $P_{\rm orb} \ge P_{\rm orb \; crit}$ {\it
the system  will remain  in the radio--ejection  phase during  all the
subsequent binary evolution}. In this case, the matter overflowing the Roche
lobe will  be accelerated  by its  interaction with  the pulsar radiation and
ejected  by the system. If we can assume (but at least ballistic simulations
should be done) that this matter leaves the system with the specific  angular
momentum at the inner Lagrangian point, $l  = (a-R_{\rm RL2}-r_1)^2 \times (2
\pi /P_{\rm orb})$ ($r_1$ is the distance between the NS and the center
of mass of the system) which in our case is quite close to the specific angular
momentum of the system, the orbital evolution is  indeed very  similar to
the  more conventional evolution with conservative mass transfer to the
primary (as confirmed by our numerical computations of the conservative case
1 and the non conservative case 2).

The optical component of \psr\ indeed seems compatible with the evolution we
have suggested. It lies along case 2 evolution, that not including
irradiation, as, in fact, the pulsar luminosity is not important as
irradiation source for the binary.

\section{Summary and conclusions} 
\label{sec:system}
We have considered the evolution of possible progenitors of the binary MSP
\psr in the Globular Cluster NGC 6397.
We can reproduce the HR diagram location of the optical companion, starting
mass transfer to the NS from a hypothetical secondary of mass 0.85 \Msun,
slightly evolved off the MS when mass transfer begins.
From equation \ref{eq:pcrit}, adopting
$\alpha =  1$, $n_{0.615} =
1$, $m_1 = 1.8$,  $m_2 = 0.45$, $\mu_{26} = 7.7$, $P_{\rm spin \; -3} = 3.65$,
and $\dot{M}_{-10} =  1$, the critical period $P_{\rm orb \; crit}$\ to reach
the `radio--ejection' phase is $\sim 39$ hr, not very different from $\simeq
P_{\rm orb}$, given the large uncertainties in the $\dot{M}_{-10}$\ estimate.
Thus, \psr is possibly in the radio--ejection phase. Given that $P_{\rm orb \;
crit}  \propto   \pspin^{9.6}~  \mu^{-4.8}~ \dot{M}^{2.04}$  the fact that
$P_{\rm orb \; crit}  \sim  P_{\rm orb}$  is compelling.

In conclusion we suggest that:
\begin{itemize}
\item[{\it i})] orbital evolution  calculations shows that a slightly
evolved 0.85 \Msun secondary orbiting a NS can transfer mass to the NS, and
reaches a stage in which its mass is reduced to $\simeq 0.45$ \Msun, and its
optical location in the HR diagram is then
compatible with the recently detected optical counterpart of \psr;
\item[{\it ii})] \psr might represent a system whose evolution has been
envisioned by Burderi et al. (2001): the spin and the magnetic moment of the
pulsar may keep the system in a radio--ejection  phase in which accretion is
inhibited by the  radiation pressure  exerted by the  pulsar on  the
overflowing matter while  the mechanism that  drives the Roche lobe  overflow
from the companion is still active, thus causing an intense wind which
would be very difficult to explain otherwise.
This evolution seems to be the only viable possibility to explain the long
lasting eclipses and the strong intensity variation randomly occurring
in the radio emission.
\end{itemize}

As a final remark we note that $P_{\rm orb} \sim P_{\rm orb \; crit}$ suggest
the interesting possibility that this system could swiftly switch from
the  present radio  pulsar phase  to an  accretion phase  in  which it
should be visible as a $L_{\rm X} \sim 10^{36}$ ergs s$^{-1}$ LMXB.

\acknowledgments{ One  of the  authors (LB) would  like to thank  T. Di
Salvo  for several  profound and  enlightening discussions  during the
preparation  of this  work.  This  work was  partially supported  by a
grant   from  the   Italian  Ministry   of  University   and  Research
(Cofin-99-02-02). We warmly thank the referee Saul Rappaport for very 
useful and stimulating criticism.}

\begin{deluxetable}{cccccccc}
\tablecolumns{7}
\tablewidth{0pc}
\tablecaption{\label{tab:dantona} Computed evolutions: $M_{1in}$=1.4\Msun,
$M_{2in}$=0.85\Msun, P$_{in}$=14.72hr}
\tablehead{
\colhead{Case} & \colhead{${M_{NS fin}}$} & \colhead{${M_{\it
WD}}$} & \colhead{${P_{orb}}$} & \colhead{AML} & \colhead{Mass transfer}
&\colhead{L$_h$} & \colhead{$\log \dot{M}_{av}$}  \\
\colhead{} & \colhead{M$_\odot$} & \colhead{M$_\odot$} & \colhead{hr} &
\colhead{} & \colhead{modalities} & \colhead{erg/s} & \colhead{M$_\odot$/yr}
}
\startdata
 1  & 2.004 & .246      & 119      & MB f=1  &  M1+M2=c &0 &-9.5   \\
    &       &           &          &         &          &  &      \\
 2  & 1.72  & .247 & 120  & MB f=1  &  M1=c from &  0 &-9.5   \\
    &       &      &      &     &  P=25.98hr &    &    \\
 3  & 2.03  & .217 & 278  & MB f=1 &  $M1+M2=c$ & 5$\times10^{34}$ &-9.9 \\
    &       &      &      &        &            &    &    \\
 4  & 1.52  & .261 & 193  & MB f=2 &  M1=c from &  0 &-9.3   \\
    &       &      &      &        &  P=17.36hr &    &    \\
\enddata
%
\end{deluxetable}


\end{document}